\documentclass[12pt]{article}
\usepackage{epsfig, amsmath, amssymb}
\usepackage{graphicx,psfrag} 
\newcommand{\be}{\begin{equation}}
\newcommand{\ee}{\end{equation}}
\newcommand{\bea}{\begin{eqnarray}}
\newcommand{\eea}{\end{eqnarray}}
\newcommand{\bA}{\begin{array}}
\newcommand{\eA}{\end{array}}
\newcommand{\bc}{\begin{center}}
\newcommand{\ec}{\end{center}}
\newcommand{\al}{\alpha}

\newcommand{\ra}{\rightarrow}
\newcommand{\del}{\partial}

\newcommand{\ie}{{\it i.e.}}
\newcommand{\eg}{{\it e.g.}}

\newcommand{\Nf}{${\cal N}{=}4$}

\begin{document}


\begin{titlepage}
\vspace{30mm}

\bc

\hfill 
\\         [22mm]

{\huge Null cosmological singularities \\ [2mm]
and free strings: II}
\vspace{16mm}

{\large K.~Narayan} \\
\vspace{3mm}
{\small \it Chennai Mathematical Institute, \\}
{\small \it SIPCOT IT Park, Padur PO, Siruseri 603103, India.\\}

\ec
\medskip
\vspace{40mm}

\begin{abstract}
In arXiv:0909.4731 [hep-th], we argued that the free string lightcone 
Schrodinger wavefunctional in the vicinity of null Kasner-like 
cosmological singularities has nonsingular time-dependence if the 
Kasner exponents satisfy certain relations. These backgrounds are 
anisotropic plane waves with singularities. We first show here that 
only certain singularities admit a Rosen-Kasner frame with exponents 
satisfying relations leading to a wavefunctional with nonsingular 
time-dependence. Then we build on the (Rosen) description further 
and study various physical observables for a time-dependent harmonic 
oscillator toy model and then the free string, reconciling this with 
the corresponding description in the conventional plane wave 
variables. We find that observables containing no time derivatives 
are identical in these variables while those with time derivatives 
are different. Various free string observables are still divergent, 
perhaps consistent with string oscillator states becoming light in 
the vicinity of the singularity. 
\end{abstract}

\end{titlepage}

{\footnotesize
\begin{tableofcontents}
\end{tableofcontents}
}

\vspace{2mm}

\section{Introduction}

In this paper, we continue exploring free string propagation in the
background of null cosmological singularities, following
\cite{knNullws2,knNullws}, and motivated by \cite{dmnt,adnt,adnnt,adgot}, 
and earlier related investigations, \eg\ 
\cite{horowitzsteif,bhkn,cornalbacosta,lms,lawrence,horopol,ckr,prt,
david,bbop,matBB,blauMpp,ohta,hertoghoro,Chu:2006pa,linwen,
Turok:2007ry,niz,niz2}.

Time-dependent generalizations of AdS/CFT were studied in
\cite{dmnt,adnt,adnnt}, the bulk containing null or spacelike
cosmological singularities (curvatures as\
$R_{MN}\sim\del_M\Phi\del_N\Phi$,\ with $e^\Phi$ a nontrivial dilaton
vanishing at the singularity), the gauge theory duals being \Nf\ Super
Yang-Mills theories with a time-dependent gauge coupling
$g_{YM}^2=e^\Phi$. In the vicinity of the bulk cosmological
singularity, supergravity breaks down with possible resolutions
arising from stringy effects (extrapolating from the usual AdS/CFT
dictionary corroborates this, with $\al'\sim {1\over g_{YM}\sqrt{N}}
\sim {1\over e^{\Phi/2}\sqrt{N}}$, suggesting vanishing effective
tension for stringy excitations as $e^\Phi\ra 0$ near the singularity). 
In particular, for the case of null singularities, the gauge theory
duals were argued to be weakly coupled \cite{dmnt}, with a
well-defined lightcone ``near-singularity'' Schrodinger wavefunctional
(barring possible renormalization effects, as discussed in
\cite{adnnt}). With a view to understanding these stringy effects but
without the complications of holography and RR-backgrounds, we studied
string propagation in purely gravitational spacetimes containing null
Kasner-like singularities in \cite{knNullws2,knNullws}. These are
essentially anisotropic plane-waves in Rosen coordinates. Building on
intuition from the Schrodinger wavefunctional as a response of the
gauge theory to a time-dependent coupling source \cite{adnnt}, we
studied the response of the free string lightcone Schrodinger
wavefunctional to the external time-dependence induced by these
cosmological singularities.  The wavefunctional was found to have
nonsingular time-dependence for certain classes of singularities
satisfying certain relations among the Kasner exponents. Furthermore, 
a detailed free string quantization can be performed using the exact
mode functions that can be solved for in these backgrounds: this shows
various string oscillator states becoming light in the vicinity of the
singularity (\eg\ on a cutoff constant-(null)time surface), relative
to the local curvature scale there.

The presence of multiple Rosen patches (\ie\ $\{A_I\}$) for a given
singularity specified by $\{\chi_I\}$ means that the conditions
$|A_I|\leq 1$ for each $I$ are nontrivial and not automatically
satisfied for a generic singularity. Indeed, using the
$\chi_I,A_I$-relations, we first analyse which singularities admit
such a Rosen frame with a nonsingular wavefunctional and find that
only certain anisotropic plane wave singularities with $\{\chi_I\}$
lying in specific windows qualify.  Since the equation of motion gives
one condition between the various $\chi_I$, the space of allowed
$\chi_I$-windows increases as the number of $\chi_I$ increases, \ie\
as the anisotropy increases.

The nonsingular time-dependence of the free string Schrodinger
wavefunctional for some singularities in the Rosen variables suggests
that string propagation might be well-defined across these
singularities in the Rosen variables. However it is essential to
understand physical observables, especially with a view to reconciling
the description in the Rosen and Brinkman frames. This is our primary
objective in this work.

We first explore a 1-dimensional time-dependent harmonic oscillator
propagating in these cosmological backgrounds in Rosen and Brinkman
variables, thinking of this as a (1-dim) single momentum mode of a
string. As for the string discussed in \cite{knNullws2}, we see that
the harmonic oscillator wavefunction for a generic state acquires a
wildly oscillating phase in Brinkman variables, suggesting that the
Rosen coordinates with a nonsingular wavefunctional above provide a
better-defined set of variables. Being a quantum mechanical system,
this can be analyzed in great detail and we study the wavefunctions,
observables and asymptotic behaviour. While the wavefunction itself is
ill-defined in the Brinkman variables but well-defined in Rosen
variables for certain values ($A\leq 1$) of the single Kasner exponent
in this model, the probability density is quite similar in both
frames, consistent with reconciling probability conservation in both
frames. While expectation values of observables not containing time
derivatives are identical in both frames as expected, those of
observables involving time derivatives are quite different: for
instance the natural momentum-squared expectation value in Rosen
variables is quite different from the corresponding Brinkman ones,
although still divergent.

This quantum mechanical analysis then paves the way for an
investigation of the free string lightcone quantization, in part
reviewing \cite{knNullws2,knNullws}. The lightcone string Schrodinger
wavefunctional has regular evolution near the singularity if the
Kasner exponents satisfy\ $|A_I|\leq 1$, for each dimension $I$. The
main new feature here is the presence of various string oscillator
states that are light near the singularity, as was argued already in
\cite{knNullws2}. This apart, various features of the observables are
similar to the harmonic oscillator case above and the expectation
values of observables containing time derivatives are different in
both frames. In a sense, this is analogous to the differences in the
variables $A_\mu$ and $\tilde A_\mu$ and various observables made from
them in the analysis of null cosmological generalizations of $AdS/CFT$
studied in \cite{dmnt}: we recall that the $A_\mu$-variables are dual
to local bulk supergravity fields which are singular, while the
$\tilde A_\mu$-variables are likely to not have local bulk duals, but
stringy ones.

Our analysis thus shows that although the wavefunctional has
nonsingular time-dependence in the Rosen variables, physical
observables in fact are still divergent for the free string, strictly
speaking, although the degree of divergence is milder than for the
Brinkman ones. It is not clear to us at this point what the
significance is, if any, of the detailed differences in the degree of
divergence of various observables in the two frames. However the
presence of light string oscillator states suggests that the free
string description is breaking down in the vicinity of the
singularity, perhaps consistent with the divergence of the free string
observables.

In sec.~2, we discuss some key features of the spacetime backgrounds
in question and the nonsingular time-dependence of the free string
lightcone Schrodinger wavefunctional, in part reviewing aspects of
\cite{knNullws2,knNullws}. In sec.~2.1, we identify conditions for the
null Rosen-Kasner exponents to admit a well-defined spacetime and
wavefunctional with nonsingular time-dependence. In sec.~3, we discuss
various aspects of the time-dependent harmonic oscillator, with a
description of the string in sec.~4. Finally sec.~5 summarises some
conclusions, with a brief discussion.

\section{Plane waves, null cosmological singularities and strings}

Consider singular plane-wave spacetimes with metric in the usual 
Brinkman coordinates
\be\label{planewave2}
ds^2=-2dy^+ dy^--\sum_I\chi_I(y^I)^2{(dy^+)^2\over (y^+)^2}+(dy^I)^2\ ,
\ee
where $\chi_I$ are real numbers characterizing the plane wave. 
As is well known, the coordinate transformation\ $y^I=(y^+)^{A_I/2} x^I ,\ 
y^-=x^-+({\sum_IA_I(y^I)^2\over 4 y^+})$, recasts these spacetimes in 
Rosen coordinate form, or manifest null cosmology form,
\be\label{Rosen1}
ds^2=-2dy^+ dx^- + (y^+)^{A_I} (dx^I)^2\ , \qquad\ 
A_I=1\pm\sqrt{1-4\chi_I}\ .
\ee
These are thus null Kasner cosmologies, the $A_I$ being Kasner exponents.\\
For this spacetime to be a solution to supergravity with no other 
background fields turned on (these in fact preserve half lightcone 
supersymmetry \cite{knNullws}), we require Ricci-flatness, \ie\
\be
R_{++} = {1\over (y^+)^2} \sum_I\chi_I = 
{1\over (y^+)^2} \sum_I {A_I (2 - A_I)\over 4} = 0\ .
\ee
Thus, completely homogenous singular plane-waves, \ie\ all $\chi_I$ equal, 
are solutions only in the presence of additional matter fields sourcing 
the system, for instance a scalar field (dilaton). For purely 
gravitational systems with no additional matter, interesting but still 
sufficiently simple spacetimes arise for just two distinct $\chi_I$, or 
$A_I$. These are then solutions if\ 
$2\chi_1+(D-4)\chi_2=2A_1(2-A_1)+(D-4)A_2(2-A_2)=0$.\ Null geodesic 
congruences stretching solely along $y^+$ (with cross-section along any 
of the other directions) have an affine parameter $y^+$\ (which can be 
seen from the geodesic equation noting that all $\Gamma^+_{ij}$ vanish).
As we approach the singularity $y^+\ra 0$, such congruences exhibit 
diverging geodesic deviation, stemming from \eg\ 
$R_{+I+I}\sim {\chi_I\over (y^+)^2}$ , so that these spacetimes exhibit 
diverging tidal forces as $y^+\ra 0$, although all curvature invariants 
are finite.

Alternative convenient forms\footnote{Indeed, we had originally found 
these in \cite{knNullws,knNullws2} as purely gravitational spacetimes 
with null scale factors $e^{f_I(x^+)}$ developing null Kasner-like 
Big-Crunch singularities $e^{f_I(x^+)}\ra (x^+)^{a_I}$ at some 
location $x^+=0$. For the case of two scale factors, one scale factor 
essentially simulates the dilaton in corresponding $AdS/CFT$ cosmology 
models \cite{dmnt,adnt,adnnt}, driving the crunch of the 4-dim part of 
the spacetime.}  of these spacetimes arise by using the 
coordinate $x^+$, with the Rosen coordinate metric
\be\label{absolns}
ds^2 = -2(x^+)^a dx^+dx^- + (x^+)^{a_I} dx^Idx^I\ ,\qquad\ \ a>0\ ,
\ee
where $I=1,2,\ldots,D-2$. Solutions with $a<0$ can be transformed to 
ones with $a>0$. The variable $y^+={(x^+)^{a+1}\over a+1}$ is the 
affine parameter for null geodesics stretched solely along $x^+$.
The corresponding plane wave Brinkman form of the metric is
\be\label{planewave}
ds^2 = -2(x^+)^a dx^+dy^- + 
\left[\sum_I \left({a_I^2\over 4}-{a_I(a+1)\over 2}\right) (y^I)^2\right] 
{(dx^+)^2\over (x^+)^2} + (dy^I)^2\ ,
\ee
with $a_I=a,b$, distinct, and $A_I={a_I\over a+1}$.
The form of the metric (\ref{absolns}) has the noteworthy feature 
that \eg\ for two Kasner exponents $a,b$, integer-values of $a,b$ exist, 
allowing for manifest analytic continuation of the metric across the 
singularity at $x^+=0$. In what follows, we will be analysing string 
propagation and we will not focus much on this feature: we will 
primarily use the metric variables in (\ref{planewave2}), (\ref{Rosen1}).

Free strings can be quantized in these backgrounds in considerable detail
since the mode functions can be exactly solved: here we review the 
analysis of \cite{knNullws2,knNullws} (which we refer to for details) 
but adapt that to the variables in (\ref{planewave2}), (\ref{Rosen1}). 
Starting with the closed string worldsheet action\ 
$S = -\int {d\tau d\sigma\over 4\pi\al'} \sqrt{-h} h^{ab}\ 
\del_a X^\mu \del_b X^\nu g_{\mu\nu}(X)$, we use lightcone gauge 
$y^+=\tau$ to reduce the system to the physical transverse degrees of
freedom. We will review some details in sec.~4.

The lightcone string worldsheet Hamiltonian in Brinkman variables is\ 
\be\label{HamilB}
H_B={1\over 4\pi\al'} \int d\sigma\ \left( (2\pi\al')^2(\Pi_y^I)^2 +
(\del_\sigma y^I)^2 +\sum_I{\chi_I\over\tau^2}(y^I)^2 \right) ,
\ee
where the conjugate momentum is\ $\Pi_y^I={\del_\tau y^I\over 2\pi\al'}$, 
which we elevate to the operator 
$\Pi^I_y[\sigma] = -i{\delta\over\delta y^I[\sigma]}$.
We see that the Hamiltonian in these variables $y^I$ contains a 
mass-term which diverges as\ $\tau\ra 0$. The wavefunctional 
$\Psi[y^I(\sigma),\tau]$ for string fields $y^I(\sigma)$ then 
acquires a ``wildly'' oscillating phase on approaching the singularity 
as $\tau\ra 0^-$,
\be
\Psi[y^I,\tau] \sim\ e^{-{i\over\tau} \sum_I\chi_I(y^I)^2}\ \Psi[y^I]\ .
\ee
Thus the wavefunction does not have a well-defined limit there.
This renders a well-defined Schrodinger wavefunctional interpretation 
near the singularity difficult in Brinkman coordinates. 

By comparison, the Rosen coordinates (\ref{Rosen1}) above naturally 
give well-defined variables for string propagation. The lightcone 
string worldsheet Hamiltonian becomes
\be\label{HamilR}
H_R = {1\over 4\pi\al'} \int d\sigma\ 
\left( (2\pi\al')^2{(\Pi^I)^2\over g_{II}} 
+ g_{II} (\del_\sigma x^I)^2 \right) = {1\over 4\pi\al'} \int d\sigma\ 
\left( (2\pi\al')^2{(\Pi^I)^2\over\tau^{A_I}} + 
\tau^{A_I} (\del_\sigma x^I)^2 \right)\ ,
\ee
with the conjugate momentum\ $\Pi^I={\tau^{A_I}\over 2\pi\al'} \del_\tau x^I$, 
which we elevate to the operator\ 
$\Pi^I[\sigma] = -i{\delta\over\delta x^I[\sigma]}$.
Now consider $A_I>0$. Then as $\tau\ra 0$, the kinetic terms 
dominate and the Schrodinger equation for the wavefunctional 
$\Psi[x^I(\sigma),\tau]$ becomes 
\be\label{SEA_I>0}
i\del_\tau\Psi[x^I,\tau] = -\pi\al' \tau^{-A_I} \int d\sigma\ 
{\delta^2\over\delta {x^I}^2}\ \Psi[x^I,\tau]
\ee
giving for the time-dependence
\be
\Psi[x^I,\tau] \sim\ e^{-i\pi\al' {\tau^{1-A_I}\over 1-A_I} \int d\sigma\
{\delta^2\over\delta {x^I}^2}}\ \Psi[x^I]\ .
\ee
The phase in the functional operator is well-defined if $A_I\leq 1$. 
Alternatively, we can recast (\ref{SEA_I>0}) as a free Schrodinger 
equation in terms of the time parameter $\tau^{1-A_I}$ (with the 
dominant exponent $A_I$) which for $A_I\leq 1$ is well-defined\ 
(vanishing as $\tau\ra 0$;\ by comparison, for $A_I>1$, this time 
parameter $\tau^{1-A_I}\ra\infty$ as $\tau\ra 0$). This free string 
Hamiltonian and associated Schrodinger equation essentially has a 
product structure with the wavefunctional itself factorizing as\ 
$\Psi=\prod_I \Psi_I[x^I]$ over the various coordinate dimensions.

For spacetimes with $A_I<0$, the potential terms dominate and we have
\be
i\del_\tau\Psi[x^I,\tau] = {\tau^{-|A_I|}\over 4\pi\al'} \int d\sigma\ 
(\del_\sigma x^I)^2 \ \Psi[x^I,\tau]\ ,
\ee
giving for the time-dependence
\be
\Psi[x^I,\tau] \sim\ e^{-i {\tau^{1-|A_I|}\over 4\pi\al'(1-|A_I|)} 
\int d\sigma\ (\del_\sigma x^I)^2} \ \Psi[x^I]\ ,
\ee
which is well-defined for $|A_I|\leq 1$. 

We see that the lightcone string Schrodinger wavefunctional for the 
generic state is well-defined near null cosmological singularities with 
Kasner exponents satisfying $|A_I|\leq 1$. A more ``nuts-and-bolts'' 
traditional quantization can be performed using the explicitly solvable 
mode functions and corroborates this, as argued in 
\cite{knNullws2,knNullws}, and discussed at length in sec.~4.

From (\ref{Rosen1}), we see that for each $\chi_I$, there exist two
Kasner exponents $A_I^{\pm}$, \ie\ two Rosen ``patches'' each. Thus we
need to be careful in identifying the precise regime where the string
wavefunctional is apparently well-defined, and in particular whether
such Rosen patches exist for a given singularity $\{\chi_I\}$ at all:
we will now discuss and elaborate on this. Then in sec.~3,
we will first discuss the time-dependent harmonic oscillator in
detail, studying the analogs of Rosen and Brinkman frames and
wavefunctions/observables therein, after which we discuss string
quantization (sec.~4), wavefunctionals and observables.

\subsection{Plane wave singularities, Rosen patches and strings}

We have seen that the Rosen frames where $|A_I|\leq 1$ encode a 
wavefunctional that has nonsingular time-dependence in the vicinity 
of the singularity. 
However since there are multiple such Rosen-Kasner exponents, it is 
not obvious if there exists a Rosen frame where $|A_I|\leq 1$ for each 
of the coordinates $x^I$. With a view to understanding this, note that 
interesting but still sufficiently simple plane wave spacetimes arise 
in the absence of matter fields (\eg\ a dilaton scalar) for just two 
distinct $\chi_I$ (and the corresponding $A_I$), as we have seen. The 
metric is\ $ds^2 = -2dy^+dx^- + \tau^{A_1} (dx_2^2+dx_3^2) + \tau^{A_2} 
(dx_4^2+\ldots+dx_{D-2}^2)$. From sec.~2, these are then solutions if\ 
$2\chi_1+(D-4)\chi_2=2A_1(2-A_1)+(D-4)A_2(2-A_2)=0$.\ 
From (\ref{Rosen1}), we see that one of the $A_I$s is positive, while
the other is negative. Indeed, given\ $\chi_I={A_I\over 4} (2-A_I)$,
the two Rosen-Kasner exponents are\ $A_I=1\pm\sqrt{1-4\chi_I}$, for
each $\chi_I$, so that $A_I^-<1$ while $A_I^+>1$ for each $I$\ (note
that $\chi_I$ is invariant under the exchange $A_I\leftrightarrow
2-A_I$, \ie\ $1-A_I\ra A_I-1$). This in all gives four Rosen frames 
or patches, $(A_1^-,A_2^-), (A_1^-,A_2^+), (A_1^+,A_2^-), (A_1^+,A_2^+)$.
\footnote{Note that for flat space, we have\ $\chi_I=0$, giving\ 
$A_I=0$ or $A_I=2$,  for each $I$. Thus the condition $|A_I|\leq 1$ 
singles out the patch\ $(A_1^-,A_2^-)$ which in this case is trivially 
flat space again. The apparent singularities in the other patches 
would seem to be coordinate artifacts: this is also possibly the case 
for the general plane wave.}

We would like a crunch happening in the ``noncompact'' directions 
$x_{2,3}$, \ie\ we impose $A_1>0$. Then the choice (with 
$\chi_1>0, \chi_2<0$)\ 
\bea
0<\chi_1\leq {1\over 4}\ , &&\quad 0<A_1=1-\sqrt{1-4\chi_1}\leq 1\ ,
\nonumber\\ 
-{3\over 4} \leq -{1\over 2(D-4)}\leq \chi_2 <0\ , && 
\quad -1\leq A_2=1-\sqrt{1-4\chi_2}<0\ ,
\eea
gives a Rosen patch $(A_1^-,A_2^-)$ which satisfies\ 
$0<|A_1|,|A_2|\leq 1$, and consistent with the equation of motion\ 
$2\chi_1+(D-4)\chi_2=0$. To elaborate, the restriction 
$\chi_1\leq {1\over 4}$ for reality of $A_1$ translates using the 
equation of motion to\ $\chi_2\geq {-1\over 2(D-4)}$, while 
$A_2\equiv A_2^-\geq -1$ (from the regularity of the Rosen 
wavefunctional) gives\ $\chi_2\geq -{3\over 4}$ : for $D>4$, these 
are compatible and satisfied if\ $\chi_2\geq {-1\over 2(D-4)}$ .
We see thus that the parameters $\chi_1,\chi_2$, must satisfy nontrivial 
conditions, lying in a particular window of parameter space: in other 
words, only certain singular plane wave spacetimes admit a Rosen patch 
of this sort. For the critical dimension $D=10$, this gives\ 
$0<\chi_1\leq {1\over 4} ,\ -{1\over 12}\leq \chi_2 <0$.

Correspondingly, we see that\ $\chi_1<0$ with $A_1>0$ requires $A_1>2$, 
which violates our requirement of regularity of the Rosen wavefunctional. 
Thus $\chi_1<0,\ \chi_2>0$, does not give any further solutions.

If we have more than 2 Kasner exponents, \ie\ more than two $\chi_I$, 
the space of possible $\chi_I$ increases, as expected. For instance, with 
say three exponents\ $A_1, A_2, A_3$, or\ $\chi_1,\chi_2,\chi_3$, arising 
from a metric 
\be
ds^2 = -2dy^+dx^- + \tau^{A_1} (dx_2^2+dx_3^2) + \tau^{A_2} (dx_4^2+dx_5^2)
+ \tau^{A_3} (dx_6^2+\ldots+dx_{D-2}^2)\ ,
\ee
we have the equation of motion\ $2\chi_1+2\chi_2+(D-6)\chi_3=0$, with\
$\chi_I={A_I(2-A_I)\over 4}$. Requiring\ $A_1>0$ as before gives\ 
$A_1=1-\sqrt{1-4\chi_1}\leq 1$: we give the various distinct allowed 
$\{\chi_I\}$ windows below ---
\begin{itemize}
\item{ $\chi_2>0, \chi_3<0$: we have\ $0<A_2=1-\sqrt{1-4\chi_2}\leq 1$ and\ 
$A_3=1-\sqrt{1+4|\chi_3|}<0$\ (note that \eg\ $A_2<0$ requires $A_2>2$ 
for $\chi_2>0$, as before, which is disallowed from the regularity of the 
Rosen wavefunctional). This is well-defined if\ 
$\chi_1,\chi_2\leq {1\over 4}$ which translates, using the equation of 
motion, to\ $-{1\over D-6}\leq \chi_3 < 0$. The regularity of the Rosen 
wavefunctional in addition requires that\ $|A_3|\leq 1$, \ie\ $A_3\geq -1$, 
\ie\ $\chi_3\geq -{3\over 4}$ as before, which is automatically satisfied 
if $D>7$.}
\item{ $\chi_2<0, \chi_3>0$: now we have\ $A_2=1-\sqrt{1+4|\chi_2|}<0$ 
and\ $0<A_3=1-\sqrt{1-4\chi_3}\leq 1$. Requiring\ 
$\chi_1,\chi_3\leq {1\over 4}$ translates using the equation of 
motion, to\ $-{D-4\over 8}\leq \chi_2<0$. Also, $A_2\geq -1$ gives as 
before\ $\chi_2\geq -{3\over 4}$: these are identical for $D=10$, while 
$-{3\over 4}\leq \chi_2<0$ is the stronger inequality if $D>10$ and 
vice versa for $D<10$.}
\item{ $\chi_2,\chi_3<0$: here we have\ $A_2=1-\sqrt{1+4|\chi_2|}<0$ 
and\ $A_3=1-\sqrt{1+4|\chi_3|}<0$. Then $A_2,A_3\geq -1$ gives\
$\chi_2,\chi_3\geq -{3\over 4}$, while\ $0<\chi_1\leq {1\over 4}$ 
translates to\ $2|\chi_2|+(D-6)|\chi_3|\leq {1\over 4}$.}
\end{itemize}

We thus see that three Kasner exponents allows several possibilities.
Clearly the story becomes richer as the spacetime becomes more 
anisotropic with multiple Kasner exponents.

Likewise, with a dilaton driving the system, there are various 
possibilities: the equation of motion in this case is\ 
$R_{++}={1\over 2} (\del_+\Phi)^2$ which simplifies to\
$\sum_I\chi_I={\al^2\over 2}$ ,\ where the dilaton time-dependence is\
$e^\Phi=t^\al$. In this case, we have the possibility of an isotropic 
crunch with all $A_I$ (or $\chi_I$) equal to say $A_I=A$: this gives\
$\chi={\al^2\over 2(D-2)}>0$, giving\ 
$0<A=1-\sqrt{1-{2\al^2\over D-2}}\leq 1$. This is well-defined if the 
dilaton exponent satisfies\ $\al\leq \sqrt{D-2\over 2}$.

We now discuss the nature of the singularity locus in these various 
patches, recalling that\ $y^I=(y^+)^{A_I/2} x^I$. Thus as\ 
$\tau\equiv y^+\ra 0$, in the Rosen null cosmology description, the 
spacetime is clearly seen to either crunch or expand depending on 
whether the corresponding Kasner exponent is $A_I>0$ or $A_I<0$. The 
singularity locus is thus the entire $\{x_I\}$-plane. In the Brinkman 
description, we see that $y^I\ra 0$ for fixed $x^I$ in the Rosen 
patch with $A_I>0$, while $y^I\ra\infty$ for $A_I<0$. Thus \eg\ we have:
\begin{itemize}
\item{Two exponents $\{\chi_1,\chi_2\}$:\ here, the singularity locus 
in Brinkman frame is $y_{2,3}\ra 0,\ y_{4,\ldots}\ra\infty$, for the 
Rosen patch $A_1>0, A_2<0$ ($\chi_1>0,\chi_2<0$), which has the 
potentially well-defined wavefunctional. This corresponds to the 
$x_{2,3}$-space crunching and the $x_{4,\ldots}$-space growing as 
$y^+\ra 0$. The other patch has\ $y_{2,3}, y_{4,\ldots}\ra 0$\ 
($A_1,A_2>0$).}
\item{Three exponents $\{\chi_1,\chi_2,\chi_3\}$: the singularity 
loci in Brinkman frame likewise are\ (i) $y_{2,3}, y_{4,5}\ra 0, 
y_{6,\ldots}\ra\infty$\ ($\chi_1,\chi_2>0, \chi_3<0$ or $A_1,A_2>0, 
A_3<0$),\ 
(ii) $y_{2,3},y_{6,\ldots}>0, y_{4,5}\ra\infty$\ ($\chi_1,\chi_3>0, 
\chi_2<0$ or $A_1,A_3>0, A_2<0$),\ 
(iii) $y_{2,3}\ra 0, y_{4,5}, y_{6,\ldots}\ra\infty$\ 
($\chi_1>0,\chi_2,\chi_3<0$ or $A_1>0, A_2,A_3<0$).} 
\end{itemize}

\section{Toy model: a time-dependent harmonic oscillator}

\subsection{Classical and preliminary quantum analysis}

We consider a 1-dim harmonic oscillator subjected to external
time-dependence: this can be regarded as an oscillator propagating in
a time-dependent background cosmological spacetime of the sort we have
discussed earlier, and is in fact a 1-dim, single momentum mode of a
string propagating in the null cosmological spacetime in Rosen
coordinates and lightcone gauge $\tau=y^+$. We will first study the
classical mechanics of this system: the (Rosen) action is
\be\label{hoscRosen}
S_R = {m\over 2} \int d\tau\ [g_{xx} {\dot x}^2 - n^2 g_{xx} x^2] = 
{m\over 2} \int d\tau\ [-2 \del_\tau x^- + \tau^A ({\dot x}^2 - n^2x^2)]\ .
\ee
The conjugate momenta are\ $p_x=m\tau^A {\dot x} ,\ 
p_-={\del L_R\over\del (\del_\tau x^-)} = -m$. The lightcone momentum 
$p_-$ is conserved. The Hamiltonian then is (using $p_-=-m$)
\be\label{HamilRosc}
H_R = p_- {\dot x^-} + p_x {\dot x} - L_R = -{p_x^2\over 2p_-\tau^A} 
- {n^2 p_-\over 2} \tau^A x^2\ .
\ee
The equations of motion then become
\be\label{dotx-}
\del_\tau(\tau^A\del_\tau x) + n^2\tau^A x = 0\ ,\qquad\
{\dot x^-} = {\del H_R\over\del p_-} = 
{1\over 2} \tau^A ( {\dot x}^2 - n^2 x^2 )\ ,
\ee
re-expressing in terms of $x,\dot x$.

Let us now consider a redefinition to Brinkman variables
\be
y=\tau^{A/2} x\ ,\qquad 
{\dot x} = \tau^{-A/2} \left( {\dot y} - {A\over 2\tau} y \right)\ ,
\ee
which have canonical kinetic terms. This, after defining
\be\label{x-y-}
y^-=x^-+{A y^2\over 4\tau}\ , \qquad \chi={A(2-A)\over 4}\ , \qquad
A = 1\pm\sqrt{1-4\chi}\ ,
\ee
then transforms the action to
\bea\label{hoscBrink}
S_B &=& {m\over 2} \int d\tau\ \left[-2 \del_\tau x^- - {d\over d\tau} 
\left({A y^2\over 2\tau}\right) + {\dot y}^2 - 
\left(n^2 + {A(2-A)\over 4\tau^2}\right) y^2\right]\nonumber\\
&=& {m\over 2} \int d\tau\ \left[-2 \del_\tau y^- + {\dot y}^2 -
\left(n^2 + {\chi\over\tau^2}\right) y^2\right]\ .
\eea
This is in fact precisely the action for a 1-dim oscillator obtained 
from a single momentum mode of the string propagating in the plane wave 
background in Brinkman coordinates.
Using this Brinkman action, we note the conjugate momenta\
$p_y=m {\dot y} ,\ 
p_-^B={\del L_B\over\del p_-} = -m = p_-^R$:\ thus the 
(conserved) lightcone momentum $p_-$ is the same in both frames. 
The Hamiltonian here is (using $p_-=-m$)
\be\label{HamilBosc}
H_B = p_- {\dot y^-} + p_y {\dot y} - L_B = -{p_y^2\over 2p_-} 
- {p_-\over 2} \left(n^2 + {\chi\over\tau^2}\right) y^2\ .
\ee
The equations of motion are
\be\label{doty-}
\ddot y + (n^2 + {\chi\over\tau^2} ) y = 0\ , \qquad
{\dot y^-} = {\del H_B\over\del p_-} = 
{1\over 2} {\dot y}^2 - {1\over 2} \left(n^2 + {\chi\over\tau^2}\right) y^2\ ,
\ee
re-expressing in terms of $y,\dot y$.

It can be seen using (\ref{dotx-}), (\ref{doty-}), that the expressions 
for $\dot x^-$ and $\dot y^-$ are consistent with (\ref{x-y-}). Starting 
with a Rosen action for just the variables $x,\dot x$ (\ie\ without 
$x^-$), the redefinition to $y,\dot y$ gives the Brinkman action 
(without $y^-$) upto a total derivative term:\ the presence of the 
variable $x^-$ in the action allows us to absorb the total derivative 
by redefining a new variable $y^-$. This procedure is thus equivalent 
to the coordinate transformation between the Rosen and Brinkman 
spacetime variables: the field redefinition automatically implements 
the coordinate transformation as it should.

We will now analyze the quantum theory of these oscillators. To begin, 
we describe some basic observations, elaborating on them later.
First we consider Brinkman variables. 
The Hamiltonian is dominated by the potential term which diverges as 
$\tau\ra 0$, for fixed momenta $p_y$ and oscillator number $n$.
The corresponding Schrodinger equation in Brinkman variables thus becomes
\be
i\del_\tau\psi_y = H_B\psi_y \sim\ -{p_-\chi\over 2\tau^2} y^2 \psi_y \qquad 
\Rightarrow\qquad i\del_{(1/\tau)} \psi \sim\ \chi y^2 \psi\ ,
\ee
which is a potential-dominated time-independent harmonic oscillator 
equation in the time variable $T={1\over\tau}$\ . However this time 
variable with $T\ra\infty$\ as\ $\tau\ra 0$\ is ill-defined: for 
instance, it takes an infinite amount of time to reach $\tau=0$ in 
the $\{T,y\}$-variables so that continuing past $\tau=0$ is not 
possible. Alternatively we can solve for the time-dependence of the 
wavefunction\ $\psi_y\sim\ e^{i{\chi\over\tau} y^2}$ , which has a wildly 
oscillating phase as $\tau\ra 0$, rendering it ill-defined there. 
This leads us to expect that the quantum system in these variables 
is ill-defined near $\tau\sim 0$. 

Now we consider Rosen variables: then for $A>0$, we see that the 
Hamiltonian $H_R$ is dominated by the kinetic term as $\tau\ra 0$. 
The Schrodinger equation becomes
\be
i\del_\tau \psi_x(\tau,x) = H_R \psi(\tau,x) \sim\ -{p_x^2\over 2p_-\tau^A} 
\psi(\tau,x) = {1\over 2p_-\tau^A} \del_x^2\psi(\tau,x)\ .
\ee
The time-dependence of the wavefunction near $\tau\ra 0$ can be solved 
for to give\\ $\psi_x(\tau,x) \sim\ e^{-i{\tau^{1-A}\over p_-} \del_x^2}\psi_x(0,x)$,
which has a well-defined phase if\ $A<1$. Alternatively, we can recast 
this as a free Schrodinger equation in terms of the variable\ 
$\tau^{1-A}\sim\lambda$. 
More precisely, consider recasting this system by redefining in 
(\ref{hoscRosen}) a new time variable given by\ 
$d\lambda=\tau^A d\tau$,\ \ie\  $\tau=((1-A) \lambda)^{1\over 1-A}$\ .
This recasts the system as a harmonic oscillator subjected solely to 
an external time-dependent frequency: the associated Schrodinger 
equation is
\be
i\del_\lambda \psi(\lambda,x) = H_\lambda \psi(\lambda,x) = 
{1\over 2} \left({p_\lambda^2\over 2p_-} + 
{n^2p_-\over 2} \lambda^{2A\over 1-A} x^2 \right) \psi(\lambda,x)\ .
\ee
As the time $\lambda$ approaches $\lambda\ra 0$, for $A<1$ the potential 
terms are subdominant and this system approaches
\be
i\del_\lambda \psi(\lambda,x) \sim {p_\lambda^2\over 2p_-} \psi(\lambda,x)\ ,
\ee
which is a free Schrodinger equation, completely regular near $\lambda=0$ 
if $A<1$. We thus choose $A=1-\sqrt{1-4\chi}$, since the other value is 
necessarily $A^+>1$. We are also assuming that $0<\chi\leq {1\over 4}$ 
for $A$ to be well-defined.

In the $\tau$-variables, although the wavefunction is regular near
$\tau=0$ for $A<1$, the Hamiltonian $H_R$ appears to be ill-defined,
with a diverging expectation value $\langle H_\tau \rangle\sim
{1\over\tau^A} \int dx \psi^* (-\del_x^2\psi)$.

This is the 1-dim harmonic oscillator analog of the string
wavefunctional described previously, with two different sets of
variables describing the physical system. While the Rosen formulation
with $H_\lambda, H_R$, appears well-defined at the level of the
wavefunctional, the Brinkman one in terms of $H_y$ appears to not
be: we will study this in greater detail in what follows.

\subsection{A more detailed analysis: wavefunctions, observables}


We will now describe in greater detail the quantization of the
harmonic oscillator in the two observer frames. There are some
parallels between this analysis and that in \cite{adnnt} for a toy
oscillator adapted from gauge theories dual to AdS cosmologies with
spacelike singularities.

In ``Brinkman'' coordinates, the equation of motion and its solution, 
\ie\ the mode function, are
\be\label{Brinkeommodes}
\ddot y + (n^2 + {\chi\over\tau^2} ) y = 0\quad\Rightarrow\quad
f_y = \sqrt{n \tau} (c_1 J_{{\sqrt{1-4\chi}\over 2}} (n\tau) + 
c_2 Y_{{\sqrt{1-4\chi}\over 2}} (n\tau))\ .
\ee
Choosing Hankel-like functions, we have early time asymptotics\ 
$f\sim e^{-in\tau}$. Quantization proceeds by taking\
\be
y = k (a_y f_y + a_y^\dag f_y^*)\ ,\qquad 
p_y = -kp_-\dot y = -p_- (a_y \dot f_y + a_y^\dag \dot f_y^*) \equiv -i\del_y\ .
\ee
The constant $k$ is fixed by the commutation relations. This gives
\be\label{ayaydag}
a_y = {\dot f_y^* y - f_y^* p_y\over k (f_y \dot f_y^* - f_y^* \dot f_y)}\ ,
\qquad
a_y^\dag = - {\dot f_y y - f_y p_y\over k (f_y \dot f_y^* - f_y^* \dot f_y)}\ ,
\ee
so that the ground state wavefunction is given by $a_y\psi_y=0$, \ie\ 
\be
p_y\psi_y(y,t) = -i\del_y \psi_y(y,t) = {\dot f_y^* \over f_y^*} y 
\psi_y (y,t)\ .
\ee
The near-singularity behaviour of this wavefunction is dictated by the 
asymptotics of the mode function\ 
$f_y\sim \lambda_1\tau^{(1+\sqrt{1-4\chi})/2}+\lambda_2\tau^{(1-\sqrt{1-4\chi})/2}$,
where the $\lambda_{1,2}$ are linear combinations\footnote{\label{Flambda}
We have\ $\lambda_1=-c_2 {n^{1-A/2}\over 2^{\sqrt{1-4\chi}/2} 
\Gamma(1+{\sqrt{1-4\chi}\over 2})} cosec(\pi{\sqrt{1-4\chi}\over 2})),\ \
\lambda_2={n^A/2\over 2^{-\sqrt{1-4\chi}/2} \Gamma(1-{\sqrt{1-4\chi}\over 2})} 
(c_1+c_2 cot(\pi({\sqrt{1-4\chi}\over 2})))$, from the Bessel 
series expansions.} of $c_{1,2}$. Using $A=1-\sqrt{1-4\chi}$, this is 
recast as\ $f_y\sim \lambda_1\tau^{1-A/2}+\lambda_2\tau^{A/2}$.
Since $A<1$, we have\ $f_y\ra c_2\tau^{A/2}$ as $\tau\ra 0$, so that 
$\psi_y(y,\tau)\sim\ e^{i {y^2\over 2\tau}}\psi(y)$, with an ill-defined 
wildly oscillating phase.
In more detail, we have\ 
$\psi_y(\tau,y) = C_y(\tau) e^{i {\dot f_y^*\over f_y^*} {y^2\over 2}}$,\ 
where the coefficient $C_y(\tau)$ is fixed by demanding that $\psi_y$ 
solve the time-dependent Schrodinger equation. Noting that $f_y$ solves 
the equation of motion, we find\ 
${\dot C_y\over C_y}=-{1\over 2} {\dot f_y^*\over f_y^*}$, 
giving the Brinkman wavefunction
\be\label{Brinkwavefn}
\psi_y(\tau,y) = {c\over \sqrt{f_y^*}}\ 
e^{i{\dot f_y^*\over f_y^*} {y^2\over 2}}\ .
\ee
Excited states constructed by acting on $\psi_y$ with $a_y^\dag$ can be 
seen to have similar singular time-dependence in the wavefunction phase.

Now let us analyze this system in the Rosen variables: the
equation of motion and the mode function solution are
\be\label{Roseneommodes}
\del_\tau(\tau^A\del_\tau x) + n^2\tau^A x = 0 \quad \Rightarrow\quad
f_x = \sqrt{n} \tau^{{1-A\over 2}} (c_1 J_{{1-A\over 2}} (n\tau) + 
c_2 Y_{{1-A\over 2}} (n\tau) )\ .
\ee
These mode functions are in fact related as\ $f_y=\tau^{A/2} f_x$, given 
the $A,\chi$-relation.
The Rosen mode function has asymptotics\ 
$f_x\sim \lambda_2 + \lambda_1\tau^{1-A}$.
These asymptotics as $\tau\ra 0$ are of course consistent with\ 
$f_y=\tau^{A/2} f_x$: this is a crucial difference between the mode 
function asymptotics in these two coordinate systems and translates to 
the striking difference between the corresponding wavefunctions.\
For what follows, we note the near singularity behaviour, 
\be\label{fasymp}
f_x\sim \lambda_2 + \lambda_1\tau^{1-A} ,\quad 
\dot f_x\sim \lambda_1 (1-A)\tau^{-A} , \quad 
f_y\sim \lambda_1\tau^{1-A/2}+\lambda_2\tau^{A/2} , \quad 
\dot f_y\sim \lambda_1(1-{A\over 2})\tau^{-A/2} + 
\lambda_2 {A\over 2} \tau^{A/2-1} ,
\ee
so that the Wronskian is\ $ f_x \dot f_x^* - f_x^* \dot f_x \sim\ 
(1-A) (c_2c_1^*-c_1c_2^*)\tau^{-A} \equiv c_0\tau^{-A}$.
With $A<1$, the asymptotics above simplifies to\ 
$f_x\sim \lambda_2 ,\ \ \dot f_x\sim \lambda_1(1-A)\tau^{-A} ,\ \ 
f_y\sim \lambda_2\tau^{A/2} ,\ \ \dot f_y\sim \lambda_2 {A\over 2} \tau^{A/2-1}$.

Quantization in the Rosen variables proceeds as
\be
x = k (a_x f_x + a_x^\dag f_x^*)\ ,\qquad 
p_x = -kp_-\tau^A \dot x = -kp_-\tau^A (a_x \dot f_x + a_x^\dag \dot f_x^*)\ .
\ee
The constant $k$ is fixed by the canonical commutation relations as
\be
[x,p_x] = i = -p_-k^2 \tau^A (f_x{\dot f_x^*}-f_x^*{\dot f_x}) [a_x,a_x^\dag] 
= -p_-k^2 c_0 \ \ \Rightarrow\ \ \ k={1\over\sqrt{|c_0 p_-|}}\ ,
\ee
where we have used the $f_x$-Wronskian ($=c_0\tau^{-A}$).
We thus have
\be\label{axaxdag}
a_x = {\dot f_x^* x - f_x^* \tau^{-A} p_x\over 
f_x \dot f_x^* - f_x^* \dot f_x}\ ,\qquad 
a_x^\dag = - {\dot f_x x - f_x \tau^{-A} p_x\over 
f_x \dot f_x^* - f_x^* \dot f_x}\ .
\ee
The ground state wavefunction then, from $a_x\psi_x=0$, is\
\be
p_x\psi_x(x,\tau) = -i\del_x\psi_x(x,\tau) = 
{\dot f_x^*\over f_x^*} \tau^A x \psi_x(x,\tau) \quad\Rightarrow\quad
\psi_x(x,\tau) \sim\ e^{i \tau^A {\dot f_x^*\over f_x} {x^2\over 2}}\ 
\ra^{\tau\ra 0}\ e^{i\tau^A {\tau^{-A}\over const}  {x^2\over 2}}\ ,
\ee
which has a well-defined phase, as $\tau\ra 0$. In more detail, the 
Rosen wavefunction is\
\be\label{Rosenwavefn}
\psi_x(\tau,x) = {c\over \sqrt{f_x^*}}\ 
e^{i\tau^A {\dot f_x^*\over f_x^*} {x^2\over 2}}\ ,
\ee
with the coefficient fixed, as for $\psi_y$, by demanding that $\psi_x$ 
solve the time-dependent Schrodinger equation.
For the case $A=0$, \ie\ no time-dependence, this wavefunction is 
the usual harmonic oscillator gaussian ground state wavefunction (with 
$f\sim e^{-in\tau}$).\ 
Excited states can be constructed by acting with $a_x^\dag$: \eg\ the 
lowest excited state can be easily seen to have the form\
$a_x^\dag \psi_x = {x\over k f_x^*} \psi_x$, using (\ref{axaxdag}), with 
the same nonsingular time-dependence in the phase of the wavefunction. 
This is also clearly true for generic excited states.

The Hilbert spaces would appear to be the same in both observer frames, 
since we have, using (\ref{ayaydag}), (\ref{axaxdag}),
\be
a_x = {1\over k (f_x \dot f_x^* - f_x^* \dot f_x)}\ 
\left(\dot f_x^* \tau^{-A/2} y - f_x^* \tau^{-A/2} 
(p_y - {A\over 2\tau} y)\right) 
= {\tau^A\over k (f_y \dot f_y^* - f_y^* \dot f_y)}\ 
( \dot f_y^* y - f_y^* p_y ) = a_y\ ,
\ee
using
\be
\dot x = \tau^{-A/2} (\dot y - {A\over 2\tau} y) \quad\Rightarrow\quad
p_x = \tau^A \dot x = \tau^{A/2} (p_y - {A\over 2\tau} y)\ ,
\ee
and\ $f_y=\tau^{A/2} f_x$, giving\ 
$f_x \dot f_x^* - f_x^* \dot f_x = \tau^{-A}(f_y \dot f_y^* - f_y^* \dot f_y)$.
\\

\noindent {\bf \emph{Observables:}}\ \ \
We will now calculate some observables in these two frames: first, we note 
using\ $f_y=\tau^{A/2} f_x ,\ 
\dot f_y = \tau^{A/2} (\dot f_x + {A\over 2\tau} f_x)$ ,\ that the Brinkman 
and Rosen wavefunctions are related as
\be\label{BRpsiReln}
\psi_y(\tau,y) = {c\over \tau^{A/4} \sqrt{f_x^*}}\ 
e^{i{(\dot f_x^*+{A\over 2\tau} f_x^*)\over f_x^*} {\tau^A x^2\over 2}}\ 
= {1\over \tau^{A/4}} e^{i\tau^{A-1} {A x^2 \over 4}}\ \psi_x(\tau,x)\ ,
\ee
the extra phase perhaps interpretable as the reflection of the canonical 
transformation ($y=\tau^{A/2} x$) between the classical Brinkman and Rosen 
variables. Note that the phase prefactor above diverging as $\tau\ra 0$ 
so that the Rosen wavefunction in Brinkman variables is apparently 
ill-defined.\\
Now we calculate the probability density: in Brinkman variables, we have
\be
|\psi_y(\tau,y)|^2 = {|c|^2\over |f_y|}\ e^{i({\dot f_y^*\over f_y^*} - 
{\dot f_y\over f_y}) {y^2\over 2}} = {|c|^2\over |f_y|}\ 
e^{i(f_y \dot f_y^* - f_y^* \dot f_y) {y^2\over 2|f_y|^2}}
= {|c|^2\over |f_y|}\ e^{-|c_0| y^2\over 2|f_y|^2}\ ,
\ee
while in Rosen variables, we have similarly
\be
|\psi_x(\tau,x)|^2 = {|c|^2\over |f_x|}\ 
e^{i\tau^A (f_x \dot f_x^* - f_x^* \dot f_x) {x^2\over 2|f_x|^2}}
= {|c|^2\over |f_x|}\ e^{-|c_0| x^2\over 2|f_x|^2}\ ,
\ee
so that
\be\label{psixy}
|\psi_y(\tau,y)|^2 = {1\over \tau^{A/2}} |\psi_x(\tau,x)|^2\ .
\ee
Using this relation (\ref{psixy}) and $y=\tau^{A/2} x$, we find on a 
constant-time surface
\be
1 = \int dy\ |\psi_y(\tau,y)|^2 = \int \tau^{A/2} dx\ 
{|\psi_x(\tau,x)|^2\over \tau^{A/2}} = \int dx\ |\psi_x(\tau,x)|^2\ ,
\ee
which is an important consistency check with probability conservation 
of our calculations in both observer frames.\\
Let us now calculate the position-squared expectation value. In Rosen 
variables, the invariant position-squared operator expectation value in 
the ground state, using (\ref{Rosenwavefn}), is
\be
g_{xx} \langle x^2 \rangle_x = \int dx\ \tau^A x^2\ |\psi_x|^2 
= \int dx\ \tau^A {x^2\over |f_x|}\ 
e^{-c_0 \tau^{-A} \tau^A {x^2\over 2|f_x|^2}} 
\sim\ \tau^A |f_x|^2\ ,
\ee
using\ $\int dx x^2 e^{-\al x^2} = {\sqrt{\pi}\over 2\al^{3/2}}$, 
while that in Brinkman variables, using (\ref{Brinkwavefn}), is
\be
\langle y^2 \rangle_y =  \int dy\ {y^2\over |f_y|}\ 
e^{-c_0 {y^2\over 2|f_y|^2}}
\equiv\ \int dx\ \tau^{A/2} {\tau^A x^2\over \tau^{A/2} |f_x|}\
e^{-c_0 {\tau^A x^2\over 2\tau^A |f_x|^2}}\ ,
\ee
which is identical to the Rosen one. Correspondingly, we have
\be
g_{xx} \langle x^2 \rangle_x = \tau^A |f_x|^2 \langle a_x a_x^\dag \rangle_0
=  |f_y|^2 \langle a_y a_y^\dag \rangle_0 = \langle y^2 \rangle_y\ .
\ee
 We see that 
$g_{xx} \langle x^2\rangle \ra 0$ as $\tau\ra 0$.

The fact that the position vevs are identical should not be surprising 
given the transformation between the variables: we expect observables 
containing derivatives might differ significantly in the two frames.
Indeed observables such as momentum-squared and correspondingly energy 
are different: \eg\ the natural Rosen frame momentum-squared expectation 
value is
\be
g^{xx} \langle p_x^2 \rangle_x = \tau^{-A} \tau^{2A} |\dot f_x|^2 \langle
a_x a_x^\dag \rangle_0 = \tau^A |\dot f_x|^2 \langle a_x a_x^\dag \rangle_0\ ,
\ee
while the natural Brinkman frame momentum-squared expectation value is
\be
\langle p_y^2 \rangle_y = \langle |\dot f_y|^2 a_y a_y^\dag \rangle_0
= \tau^A |\dot f_x + {A\over 2\tau} f_x|^2 \langle a_x a_x^\dag \rangle_0
\ee
which are quite different. In particular in the vicinity of the 
singularity $\tau\ra 0$, the asymptotics are
\be\label{oscMomvevAsymp}
Rosen:\ \ \ \sim\ \tau^{-A} \langle a_x a_x^\dag \rangle_0\ ,
\qquad\ 
Brinkman:\ \ \ \sim\ \tau^{A-2} \langle a_x a_x^\dag \rangle_0\ .
\ee
For $A<1$, as $\tau\ra 0$, we see that the Brinkman momentum-squared 
is more divergent. Note however that there exist a set of Brinkman 
observables that simulates the natural Rosen observables.

The expressions (\ref{HamilRosc}), (\ref{HamilBosc}), for the Hamiltonian in 
the Rosen and Brinkman frames can be simplified to obtain
\be
H_R = -{k^2p_-\tau^A\over 2} \left[ a_x^2 (\dot f_x^2 + n^2 f_x^2) +
(a_x^\dag)^2 (\dot {f_x^*}^2 + n^2 {f_x^*}^2) + 
(a_xa_x^\dag+a_x^\dag a_x) (|\dot f_x|^2 + n^2 |f_x|^2) \right]\ ,
\ee
and
\bea
H_B &=& -{k^2p_-\over 2} \Bigl[ (a_ya_y^\dag+a_y^\dag a_y) 
(|\dot f_y|^2 + (n^2 + {\chi\over\tau^2}) |f_y|^2) \nonumber\\ 
&& \qquad\qquad\qquad\ + a_y^2 (\dot f_y^2 + (n^2 + {\chi\over\tau^2}) f_y^2)
+ (a_y^\dag)^2 (\dot {f_y^*}^2 + (n^2 + {\chi\over\tau^2}) {f_y^*}^2) \Bigr]\ .
\eea
Note that the Rosen and Brinkman Hamiltonians do not transform into 
each other through the field redefinitions: \eg\ the coefficients of 
$\{a,a^\dag\}$ in the two expressions differ as\
$\tau^A (|{\dot f_x}|^2 + n^2 |f_x|^2) =
|{\dot f_y}|^2 + (n^2-{\chi\over\tau^2}) |f_y|^2 
- {d\over d\tau} ({A |f_y|^2\over 2\tau})$ , consistent with the 
fact that the Lagrangians differ by the total derivative term.
 The ground state energy expectation values are
\be
\langle H_R\rangle \sim\ \tau^A \tau^{-2A} k^2|p_-|\ ,\qquad
\langle H_B\rangle \sim\ \tau^{A-2} k^2|p_-|\ .
\ee

To see this in dimensionful detail, we note that\ 
$dim [x^+]=dim [\tau]=dim [x^-]=length (L),\\ dim [(x^+)^{A/2} x]=dim [y]=L,
\ dim [m]=L^{-1},\ dim [p_-]=L^{-1}$.\ Then the potential term in 
(\ref{hoscRosen})
(and correspondingly in (\ref{hoscBrink})) is really ${n^2\over l^2} x^2$ , 
where $l$ is the string coordinate length with $dim [l]=L$. This gives 
\eg\ the dimensionless modes\ $f_x = \sqrt{n} ({\tau\over l})^{{1-A\over 2}} 
(c_1 J_{{1-A\over 2}} ({n\tau\over l}) + c_2 Y_{{1-A\over 2}} ({n\tau\over l}))$,
and\ $[x,p_x]=i={p_-c_0 k^2\over (1-A) l^{1-A}} [a_x,a_x^\dag]$, so that\
$k=\sqrt{l^{1-A}\over |c_0 p_-|}$ , which has\ $dim [k]=dim [x]$. Finally 
this gives\ $H_R\sim |p_-| {n^2\over l} {l^A l^{1-A}\over |p_-|} 
(a_x^\dag a_x + \ldots)$, so that\ $dim [H_R]=dim [{1\over l}]$.

\subsection{The free particle}

This is essentially the $n=0$ limit of the harmonic oscillator we have 
already discussed: however it is instructive.
So consider the free particle with (Rosen) action, momenta and Hamiltonian
\be
S_R = {m\over 2} \int d\tau\ [-2 \del_\tau x^- + \tau^A {\dot x}^2]\ , 
\qquad H_R = -{p_x^2\over 2p_-\tau^A}\ , \qquad 
p_-={\del L_R\over\del (\del_\tau x^-)} = -m ,\quad p_x=m\tau^A \dot x\ .
\ee
Then the equation of motion is\ $\del_\tau (\tau^A\del_\tau x)=0$,\ giving\ 
$x=c_1\tau^{1-A}+c_2$ and ${\dot x^-} = {\del H_R\over\del p_-} = 
{1\over 2} \tau^A {\dot x}^2 \sim c_1 \tau^{-A}$, \ie\ $x^-\sim \tau^{1-A}$. 
If $A<1$, these are well-defined particle trajectories, with finite 
$x\ra c_2=const$, \ie\ non-diverging as $\tau\ra 0$. The Schrodinger 
time evolution is well-defined for fixed $x, p_x$.

Now consider the redefinition to the Brinkman variable\ $y=\tau^{A/2} x$. 
We have\ $\dot x=\tau^{-A/2} (\dot y - {A\over 2\tau} y)$, giving\
\be
S_B={m\over 2} \int d\tau\ \left[-2 \del_\tau y^- + {\dot y}^2 -
{\chi\over\tau^2} y^2\right]\ ,\qquad\
H_B = -{p_y^2\over 2p_-} - {p_-\chi\over 2\tau^2} y^2\ ,
\ee
where we have used the conjugate momenta\ $p_y=m {\dot y} ,\ 
p_-^B={\del L_B\over\del p_-} = -m = p_-^R$.
For this Brinkman $S_B$ system, we have the equation of motion\ 
$\del_\tau^2 y + {\chi\over\tau^2} y = 0$,\ giving the solutions\ 
$y=c_1\tau^{1-A/2}+c_2\tau^{A/2}=\tau^{A/2} x$, where\ $A=1-\sqrt{1-4\chi}$.
These trajectories all have\ $y\ra 0$ as $\tau\ra 0$.
Also we have\ ${\dot y^-} = {1\over 2} {\dot y}^2 - {1\over 2} 
{\chi\over\tau^2} y^2 \sim \tau^{A-2}$, giving\ $y^-\sim \tau^{A-1}$, 
which diverges.

While the Brinkman system looks like a time-dependent harmonic
oscillator with a divergent frequency ${\chi\over\tau^2}$, the
classical solutions are not oscillatory. 
The Hamiltonians here are in fact those obtained from free particle
propagation seen by Rosen and Brinkman observers respectively.
The Schrodinger time evolution is ill-defined for fixed $y, p_y$.

\section{String quantization: Rosen and Brinkman variables}

We now continue our discussion of string propagation following from 
sec.~2 earlier, where we have described the Hamiltonians and the 
lightcone string Schrodinger wavefunctionals in the two frames. We saw
there that the lightcone string Schrodinger wavefunctional for the 
generic state has well-defined evolution in the vicinity of the 
singularity in Rosen variables if the Kasner exponents satisfy 
$|A_I|\leq 1$. We will now discuss string quantization in greater 
detail in Rosen and Brinkman variables: in part this will review the 
analysis of \cite{knNullws2,knNullws}. There are close parallels with 
our earlier discussion on the time-dependent harmonic oscillator 
which is a 1-dim single momentum mode of the string. 

The closed string worldsheet action\ 
$S = -\int {d\tau d\sigma\over 4\pi\al'} \sqrt{-h} h^{ab}\ 
\del_a X^\mu \del_b X^\nu g_{\mu\nu}(X)$ takes the form\
$S_R = -{1\over 4\pi\al'} \int d^2\sigma\ (
- 2E g_{+-} \del_\tau X^- - E g_{II} (\del_\tau x^I)^2 
+ {1\over E}\ g_{II} (\del_\sigma x^I)^2)$, in Rosen variables,
upon using lightcone gauge $y^+=\tau$ and setting $h_{\tau\sigma}=0$, 
$E(\tau,\sigma)=\sqrt{-{h_{\sigma\sigma}\over h_{\tau\tau}}}$ , as in 
\cite{Polchinski:2001ju} (see also \cite{Metsaev:2000yf}). 
The momentum conjugate to $x^-$ can be fixed to a $\tau$-independent 
constant\ $p_-={E g_{+-}\over 2\pi\al' l}=-{1\over 2\pi\al' l}$ by a 
$\tau$-independent $\sigma$-reparametrization invariance (not fixed by 
the gauge fixing above), giving\ \ $E=-{1\over g_{+-}}=1$. The quantity 
$l$ here is the string coordinate length, which from above becomes\
$l=-2\pi p_-\al'=2\pi|p_-|\al'$ \ (note that $p_-\leq 0$ always, and 
we will sometimes use $p_-$ to denote $|p_-|$). [Our conventions here 
agree with \cite{joetext} for flat space.]
Thus we see that for the spacetime backgrounds we consider (with 
$g_{+-}=-1$), lightcone gauge $y^+=\tau$ is compatible with conformal 
gauge\ $h_{ab}=\eta_{ab}$. 
The string Hamiltonian, $H=-p_+$, in Rosen variables (\ref{HamilR}), 
after re-expressing the momenta in terms of derivatives, becomes
\be
H_R={1\over 4\pi\al'} \int_0^l d\sigma\ \tau^{A_I} \left( (\del_\tau x^I)^2
+ (\del_\sigma x^I)^2 \right)\ ,
\ee
containing only the physical transverse modes $x^I(\tau,\sigma)$. The 
range of $\int d\sigma$ is $\int_0^{2\pi|p_-|\al'} d\sigma$, 
involving the lightcone momentum $p_-$. This Hamiltonian is the 
physical Hamiltonian\ $H=-p_+$ satisfying the physical state condition\ 
$m^2=-2g^{+-}p_+p_--g^{II}(p_{I0})^2$. In lightcone gauge $y^+=\tau$, 
the worldsheet Schrodinger equation also essentially governs the 
evolution in spacetime $i{\del\over\del y^+} \Psi$ of the string 
wavefunctional. From the Hamiltonian $H[x^-,p^-,x^I,\Pi^I]$, we can 
solve for $x^-$ using $\del_\tau x^-={\del H\over\del p_-}$.

Similarly, the string action in Brinkman variables is\
$S_B = -{1\over 4\pi\al'} \int d^2\sigma\ ( 2 \del_\tau y^- - 
(\del_\tau y^I)^2 + (\del_\sigma y^I)^2 + {\chi_I\over \tau^2} (y^I)^2)$ ,
with the Brinkman Hamiltonian (\ref{HamilB}) becoming
\be
H_B={1\over 4\pi\al'} \int_0^l d\sigma\ \left( (\del_\tau y^I)^2 +
(\del_\sigma y^I)^2 +{\chi_I\over\tau^2}(y^I)^2 \right)\ .
\ee

As should be clear, single oscillator/momentum modes of the string are
essentially identical in description to the time-dependent harmonic
oscillator discussed earlier. In what follows, we will discuss the 
explicit quantization of the string.

The mode expansion for the spacetime coordinate fields of the string is
\be\label{modeexpXIn}
X^I(\tau,\sigma) = X^I_0(\tau) + \sum_{n=1}^\infty \left( 
k_n^I f^I_n(\tau) (a^I_n e^{in\sigma/l} + {\tilde a}^I_n e^{-in\sigma/l}) + 
k_n^{I*} f^{I*}_n(\tau) (a^I_{-n} e^{-in\sigma/l} + {\tilde a}^I_{-n} 
e^{in\sigma/l}) \right)\ ,
\ee
where $f^I_n(\tau)$ are the mode functions in either Rosen or Brinkman 
coordinate frames, following from the worldsheet equations of motion. 
The conjugate momentum is\ $\Pi^I={\tau^{A_I}\over 2\pi\al'} \del_\tau x^I$ 
and $\Pi_y^I={\del_\tau y^I\over 2\pi\al'}$ in Rosen and Brinkman 
variables respectively.
The constants $k^I_n$ are fixed as\ 
$k_{Bn}^I={i\over n}\sqrt{{\pi\al'\over 2|c^I_{n0}|}}$ (Brinkman variables) 
and\ $k_{Rn}^I={i\over n}\sqrt{{\pi\al' l^{-A_I}\over 2|c^I_{n0}|}}$ (Rosen 
variables) by demanding that the canonical commutation relations for the 
fields be consistent with the creation-annihilation operator algebra\
$[a^I_n,a^J_{-m}]=[{\tilde a}^I_n,{\tilde a}^J_{-m}]=n\delta^{IJ}\delta_{nm}$ 
(or equivalently, the canonical commutation relations for the $X^I,\Pi^I$).
This effectively makes the mode functions $f^I_n$ dimensionless. 
The no-scale property of these spacetimes, \ie\ requiring no explicit 
length scale, is manifest in the Brinkman form, so that the dimensions 
of the Rosen variables $x^I$ are nontrivial for consistency. Nontrivial 
factors of $l$ enter in the string quantization in accord with this.

To calculate the string Hamiltonian, let us first use the mode
expansion (\ref{modeexpXIn}) to evaluate
\bea\label{intermediateterms}
{1\over l} \int_0^l d\sigma (\del_\tau X^I)^2 &=& ({\dot X}^I_0)^2 
+ \sum_n  |k^I_n|^2 \Big( |{\dot f}^I_n|^2 ( \{a^I_n,a^I_{-n}\} + 
\{{\tilde a}^I_n,{\tilde a}^I_{-n}\} ) 
- ({\dot f}^I_n)^2 \{a^I_n,{\tilde a}^I_n\} \nonumber\\
&& {} \qquad\qquad\qquad\ 
- ({\dot f}^{I*}_n)^2 \{a^I_{-n},{\tilde a}^I_{-n}\} \Big)\ , \nonumber\\
{1\over l} \int_0^l d\sigma (\del_\sigma X^I)^2 &=&
\sum_n n^2 |k^I_n|^2 \Big( |f^I_n|^2 ( \{a^I_n,a^I_{-n}\} + 
\{{\tilde a}^I_n,{\tilde a}^I_{-n}\} ) 
- (f^I_n)^2 \{a^I_n,{\tilde a}^I_n\}  \nonumber\\
&& {} \qquad\qquad\qquad\qquad\qquad
- (f^{I*}_n)^2 \{a^I_{-n},{\tilde a}^I_{-n}\} \Big)\ ,\nonumber\\
{1\over l} \int_0^l d\sigma (X^I)^2 &=& 
(X^I_0)^2 + \sum_n |k^I_n|^2 \Big( |f^I_n|^2 ( \{a^I_n,a^I_{-n}\} + 
\{{\tilde a}^I_n,{\tilde a}^I_{-n}\} ) 
- (f^I_n)^2 \{a^I_n,{\tilde a}^I_n\}  \nonumber\\
&& {} \qquad\qquad\qquad\qquad\qquad
- (f^{I*}_n)^2 \{a^I_{-n},{\tilde a}^I_{-n}\} \Big)\ .
\eea

\noindent The Rosen Hamiltonian then simplifies to
\bea\label{HamilRadaga}
H_R &=& {l\over 2\al'} \tau^{A_I} {({\dot x}^I_0)^2\over 2} 
 +\ {l\over 2\al'} \sum_n |k^I_{Rn}|^2 \tau^{A_I} \Biggl( ( \{a^I_n,a^I_{-n}\}
+ \{{\tilde a}^I_n,{\tilde a}^I_{-n}\} ) \left( |{\dot f}^I_{Rn}|^2 + 
{n^2\over l^2} |f^I_{Rn}|^2 \right) \nonumber\\
&& \qquad\quad\ -\ \{a^I_n,{\tilde a}^I_n\} \left( ({\dot f}^I_{Rn})^2 + 
{n^2\over l^2} (f^I_{Rn})^2 \right) 
- \{a^I_{-n},{\tilde a}^I_{-n}\} \left( ({\dot f}^{I*}_{Rn})^2 
+ {n^2\over l^2} (f^{I*}_{Rn})^2 \right) \Biggr) ,
\eea
while the Brinkman Hamiltonian becomes
\bea\label{HamilBadaga}
H_B &=& {l\over 2\al'} \left({({\dot y}^I_0)^2\over 2} + {\chi\over\tau^2} 
(y^I_0)^2\right)\ + \nonumber\\
&&  {l\over 2\al'} \sum_n |k^I_{Bn}|^2 \Biggl[ ( \{a^I_n,a^I_{-n}\} + 
\{{\tilde a}^I_n,{\tilde a}^I_{-n}\} ) \left( |{\dot f}^I_{Bn}|^2 + 
({n^2\over l^2} + \sum_I{\chi_I\over\tau^2}) |f^I_{Bn}|^2 \right)\ - \\
&&  \{a^I_n,{\tilde a}^I_n\} \left( ({\dot f}^I_{Bn})^2 + 
({n^2\over l^2} + \sum_I{\chi_I\over\tau^2}) (f^I_{Bn})^2 \right) 
- \{a^I_{-n},{\tilde a}^I_{-n}\} \left( ({\dot f}^{I*}_{Bn})^2 
+ ({n^2\over l^2} + \sum_I{\chi_I\over\tau^2}) (f^{I*}_{Bn})^2 \right) \Biggr]
 .\nonumber
\eea
The mode functions in the mode expansion above are
\bea
R:\qquad f^I_{Rn}(\tau) &=& {\sqrt{n}\tau^{1-A_I\over 2}\over l^{1-A_I\over 2}}
(c^I_{n1} J_{{1-A_I\over 2}}({n\tau\over l}) + 
c^I_{n2}Y_{{1-A_I\over 2}}({n\tau\over l}))\ ,
\nonumber\\
B:\qquad
f^I_{Bn}(\tau) &=& {\sqrt{n\tau}\over \sqrt{l}} 
(c^I_{n1} J_{{\sqrt{1-4\chi_I}\over 2}}({n\tau\over l})
+ c^I_{n2} Y_{{\sqrt{1-4\chi_I}\over 2}}({n\tau\over l}))\ ,
\eea
which, using the relation (\ref{Rosen1}) between $A_I,\chi_I$, are
related as\ $f^I_{Bn}=({\tau\over l})^{A_I/2} f^I_{Rn}$, and
correspondingly for the asymptotic forms too. Using this relation
between the modes, we see that the Rosen and Brinkman Hamiltonians do
not transform into each other through the field redefinitions: \eg\
the coefficient of $\{a^I_n,a^I_{-n}\}$ in the Rosen Hamiltonian is
\be
{|k^I_{Rn}|^2\over |k^I_{Bn}|^2} \tau^{A_I} \left(|{\dot f}^I_{Rn}|^2 + 
{n^2\over l^2} |f^I_{Rn}|^2\right) =
|{\dot f}^I_{Bn}|^2 + \left({n^2\over l^2}-{\chi_I\over\tau^2}\right) 
|f^I_{Bn}|^2 - {d\over d\tau} \left({A_I |f^I_{Bn}|^2\over 2\tau}\right) .
\ee
This is consistent with the fact that the field redefinition 
$y^I=\tau^{A_I/2} x^I$ transforms $x^-, y^-$ in the corresponding Rosen 
and Brinkman Lagrangians by the total time derivative term.

The string oscillator masses are then given by\
$m^2=-2g^{+-} H p_--g^{II} (p_{I0})^2$, which simplifies to the terms 
containing the oscillator operators in the Hamiltonians. Thus we see 
that the detailed string spectra in the Rosen and Brinkman frames are 
different.

To find the mode asymptotics in the vicinity of the singularity, define 
a cutoff surface at constant $y^+\equiv\tau=\tau_c$ a little away from 
the singularity at $y^+=0$.
Then the low lying states (finite $n\ll {1\over \tau_c}$) have mode 
asymptotics essentially identical to the harmonic oscillator described 
earlier, with power law behaviour\ $f^I_{Rn}\sim 
\lambda^I_{2n} + \lambda^I_{1n}({\tau_c\over l})^{1-A_I}$, with corresponding 
asymptotics for the Brinkman modes\ (the $\lambda^I_n$s arise from the 
Bessel series expansions; see footnote~\ref{Flambda} in the harmonic 
oscillator case). A single low lying oscillator 
mode-$\{I,n\}$ ($n\ll {1\over\tau_c}$) has mass\ 
$m^2\sim\ g^{+-}H_Rp_- \sim\ {l^{A_I}\over\al'\tau_c^{A_I}}$ in Rosen 
variables,\ which is light relative to the local curvature scale 
${1\over\tau_c^2}$ , if\ ${l^{A_I}\over\al'\tau_c^{A_I}}\ll {1\over\tau_c^2}$.
Likewise in Brinkman variables, a single low lying oscillator mode 
has mass $m^2\sim g^{+-}H_Bp_-\sim\ {l^{1-A_I}\over\tau_c^{2-A_I}}$ ,\
different in form from the Rosen one: this is light relative to the 
local curvature scale if\ ${\tau_c^{A_I} p_-l\over l^{A_I}}\ll 1$. These 
two conditions look a priori different: they are consistent if\ 
${\tau_c^2\over\al'} \ll ({\tau_c\over l})^{A_I} \ll {1\over p_-l}$, 
which using $l\sim p_-\al'$, simplifies to\ 
$p_-\ll {1\over\tau_c}\equiv {1\over y^+_c}$. This condition also 
arises from consistency for other modes being light as we will see later.

We now see that for any cutoff $\tau_c$ no matter how small, there 
exist highly stringy modes defined by the limit of small $\tau_c$, 
large $n$, with $n\tau_c\gg 1$. These are modes that effectively do 
not see the singularity, thus behaving like flat space modes. In the 
Rosen frame, these modes have\
$f^I_{Rn}\sim\ {e^{-in\tau_c/l}\over (\tau_c/l)^{A_I/2}}$, for 
$c^I_{n1}=1, c^I_{n2}=-i$, and the Hamiltonian simplifies to
\be\label{HRstringy}
H_R\sim\ {1\over l} \sum_{n\gg 1/\tau_c} 
(a^I_{-n}a^I_n+{\tilde a}^I_{-n}{\tilde a}^I_n+n)\ .
\ee
The corresponding oscillator mass for a single highly stringy mode is 
(using $l=2\pi|p_-|\al'$)
\be
m^2 \sim\ -g^{+-} H_R p_- \sim\ {n\over\al'}\ .
\ee

Similarly in the Brinkman frame, we have\ 
$f^I_{Bn}\sim\ e^{-in\tau_c/l}$ for $c^I_{n1}=1, c^I_{n2}=-i$, with the 
Hamiltonian simplifying to
\be\label{HBstringy}
H_B \sim\ {1\over l} \sum_{n\gg 1/\tau_c}\ (1-{\chi_I\over 2n^2\tau_c^2}) 
(a^I_{-n}a^I_n+{\tilde a}^I_{-n}{\tilde a}^I_n+n) 
- {\cal O}({1\over (n\tau_c)^2})\ ,
\ee
the same as (\ref{HRstringy}) to leading order, and the mass of a 
single highly stringy mode becoming\
$m^2 \sim\ -g^{+-} H_B p_- \sim\ {n\over\al'}$.
Note that the norms of the mode amplitudes are\ $g_{II} |f^I|^2$,
which is the same in both observer frames, \ie\ 
$\tau^A ({1\over \tau^{A/2}})^2\sim\ const$ (Rosen) and likewise for 
Brinkman.
We see that the highly stringy modes have similar structure in either 
frame, perhaps not surprisingly since these are essentially modes that 
are sufficiently high frequency that they effectively do not see the 
approaching singularity at the hypersurface $\tau=\tau_c$. Now  
note that in the vicinity of the cutoff surface $y^+=y^+_c$, the local 
curvature scale is\ ${1\over (y_c^+)^2}$. Thus various highly stringy 
single oscillator states are light on a near-singularity cutoff 
surface if\ $m^2\ll {1\over (y^+_c)^2}$. This gives
\be\label{lightosc}
{p_-\al'\over (y^+_c)} \ll n \ll {\al'\over (y^+_c)^2}\ ,
\ee
the first inequality arising from our definition of highly stringy 
modes. This implicitly implies $p_-\ll {1\over y^+_c}$. Note that the 
coordinate length $l=2\pi|p_-|\al'$ of the string increases as the 
lightcone momentum $p_-$ increases. Thus we have\ 
$l\ll l_s ({l_s\over y^+_c})$. For a Planck scale cutoff $y^+_c\sim l_P$, 
we thus have $l\ll {l_s\over g_s^{2/(D-2)}}$, using the naive relation
for the Newton constant $G_D=l_P^{D-2}=g_s^2 l_s^{D-2}$. Thus in the 
weakly coupled (or free) string limit, the string coordinate length 
can be large for these states in the vicinity of the singularity, 
\ie\ the string can effectively be a large floppy object.
The number of such oscillator levels excited is\ 
${\al'\over (y^+_c)^2} (1-p_-y^+_c)$. In the singular limit $y^+_c\ra 0$, 
all oscillator states are light and the number of excited oscillator 
states diverges. Conversely in the sector\ $p_-\sim {1\over y^+_c}$, the 
window of light highly stringy states pinches off. On a string scale 
cutoff surface $y^+_c\sim l_s$, we see that no string oscillators are 
turned on, \ie\ $n\sim 1$ is already not a light state from 
(\ref{lightosc}). On a Planck scale cutoff surface, the highest 
oscillator level turned on is of order\ 
$n\sim ({l_s\over l_p})^2\sim {1\over g_s^{2/(D-2)}}$, \ie\ in the free 
string limit $g_s\ra 0$, we have a large number $n\gg 1$ of highly 
stringy oscillator states \cite{knNullws2}.

Thus various (highly stringy) light oscillator states arise near the 
singularity, in all observer frames, suggesting string interactions 
are non-negligible near the singularity. \\

\noindent {\bf \emph{Observables:}}\ \
Let us now calculate the expectation value of the position-coordinate-squared 
for the string. Considering the string to be a discretized set of 
oscillators, this would be\ $\langle \sum_k g_{II} (X^I_{\sigma_k})^2 \rangle$, 
which in the continuum limit becomes \
$\langle {1\over l}\int d\sigma\ g_{II} (X^I)^2\rangle$. 
Since different coordinate directions behave differently due to the 
anisotropy of the spacetime, we calculate this expectation value for each 
direction $I$ separately, \ie\ the index $I$ is not summed over.
We then have, using (\ref{intermediateterms}),
\be
\langle {1\over l}\int d\sigma\ g_{II} (X^I)^2\rangle 
= (X^I_0)^2 + \sum_n |k^I_n|^2 g_{II} |f^I_n|^2 ( \{a^I_n,a^I_{-n}\} + 
\{{\tilde a}^I_n,{\tilde a}^I_{-n}\} )\ ,
\ee
where we are considering states with single excitations, \eg\ 
$a^I_{-m} {\tilde a}^J_{-m} |0\rangle$. It is clear that this observable 
is identical in both Rosen and Brinkman frames, as for the harmonic 
oscillator discussed previously, noting that the mode functions are 
related as\ $f^I_{Bn}=({\tau\over l})^{A_I/2}f^I_{Rn}$.
In the Rosen variables, since $f^I\ra constant$ for the low lying
oscillator modes, we see that the spatial directions with $A_I>0$ have
the string shrinking, while those with $A_I<0$ have the string
elongating. For the highly stringy oscillator modes, we have\ 
$f^I\ra {e^{-in\tau}\over\tau^{A_I}}$, so that\ $\langle {1\over l}\int
d\sigma\ g_{II} (X^I)^2\rangle \ra\ const$.

It is natural to define the momentum-squared expectation value likewise
as the continuum version of the discretized observable
$\langle \sum_k g^{II} (\Pi^I_{\sigma_k})^2 \rangle = 
\langle \sum_k g_{II} (\del_\tau X^I_{\sigma_k})^2 \rangle$, which becomes\
$\langle {1\over l}\int d\sigma\ g^{II} (\Pi^I)^2\rangle = 
\langle {1\over l}\int d\sigma\ g_{II} (\del_\tau X^I)^2\rangle$.
Then, using (\ref{intermediateterms}), we have
\be
\langle {1\over l}\int d\sigma\ g^{II} (\Pi^I)^2 \rangle 
= g_{II} (\dot X^I_0)^2 + \sum_n |k^I_n|^2 g_{II} |\dot f^I_n|^2 
( \{a^I_n,a^I_{-n}\} + \{{\tilde a}^I_n,{\tilde a}^I_{-n}\} )\ ,
\ee
again considering states with single excitations, \eg\ 
$a^I_{-m} {\tilde a}^J_{-m} |0\rangle$.\ We see now that the natural 
momentum observable in the Rosen frame is\ 
$\langle {1\over l} \int d\sigma\ \tau^{A_I} (\del_\tau x^I)^2\rangle$, 
whereas it is $\langle {1\over l} \int d\sigma\ (\del_\tau y^I)^2\rangle$ 
in the Brinkman frame. These however are not the same clearly: we have
\be
{(\Pi^I_x)^2\over\tau^{A_I}} = \left(\Pi^I_y - {A_I\over 2\tau} y^I\right)^2\ .
\ee
Part of the kinetic energy of the string in the Rosen frame looks 
like a potential energy in the Brinkman frame with a time-dependent 
prefactor that diverges near the singularity. In detail, we see from 
the asymptotics of the mode functions $f^I_{Rn}, f^I_{Bn}$,\ (similar to 
(\ref{fasymp}) for the time-dependent harmonic oscillator) that the 
Rosen momentum-squared expectation value is less divergent than the 
Brinkman one (for $|A_I|\leq 1$)
\bea
Rosen:&& \quad \langle{1\over l}\int d\sigma\ \tau^{-A_I} (\Pi_x^I)^2\rangle 
\sim\ \tau^{A_I} |\dot f^I_{Rn}|^2 \sim\ \tau^{-A_I}\ ,\nonumber\\
Brinkman:&& \quad \langle{1\over l}\int d\sigma\ (\Pi_y^I)^2\rangle \sim\ 
|\dot f^I_{Bn}|^2 \sim\ \tau^{A_I-2}\ ,
\eea
very similar to (\ref{oscMomvevAsymp}) for the harmonic oscillator.

The expectation value of the energy can be evaluated in similar
fashion. Focussing on the singularities with $|A_I|\leq 1$ for each $I$ 
(in the next section, we show that such singularities exist but only 
in certain windows for the $\{\chi_I\}$), the mode function asymptotics 
as $\tau\ra 0$ are\ $f^I_{Rn}\sim\lambda^I_{2n},\ \dot f^I_{Rn}\sim 
{\lambda^I_{1n}\over l} ({\tau\over l})^{-A_I}$. Since the time-dependent 
factor (in the Rosen variables) in the Hamiltonian appears as 
$\tau^{\pm A_I}$ in the kinetic and potential energy terms respectively, 
for both $A_I>0$ and $A_I<0$ one factor will dominate as $\tau\ra 0$ 
and diverge as $\tau^{-|A_I|}$. Thus the Hamiltonian expectation value 
in a state with some oscillators in Rosen variables, using 
(\ref{HamilRadaga}) and (\ref{HRstringy}), on a cutoff surface 
$\tau=\tau_c$, is
\be
\langle H_R\rangle \sim\ {l\over 2\al'} \tau_c^{A_I} {({\dot x}^I_0)^2\over 2}
+\ {l\over 2\al'} \sum_{n\ll {1\over\tau_c}} {|k^I_{Rn}|^2\over\tau_c^{|A_I|}} 
( N^I_n + {\tilde N}^I_n + n) C^I_n +\ 
{1\over l}\sum_{n\gg {1\over\tau_c}} (N^I_n+{\tilde N}^I_n + n)\ ,
\ee
where the constants $C^I_n$ are either ${n^2\over l^2} |\lambda^I_{2n}|^2$
or ${|\lambda^I_{1n}|^2\over l^2}$ depending on whether $A_I<0$ or $A_I>0$ 
respectively, the $N^I_n, {\tilde N}^I_n$ being the number of 
oscillators turned on. We have approximated the energy contributions 
from the low lying and highly stringy states on the cutoff surface. 
Since the number of such oscillator levels (both low lying and highly 
stringy) excited increases as $y^+\ra 0$, we expect that the total 
energy imparted to the string also increases, and the expression above 
indicates a divergence as we approach the singularity.\\
The Brinkman Hamiltonian expectation value $\langle H_B\rangle$ has 
similar structure, except with the zero mode contribution being\ 
${l\over 2\al'} ({({\dot y}^I_0)^2\over 2} + {\chi\over\tau_c^2} (y^I_0)^2)$ ,
\ and the low lying oscillator contribution having a time-dependent 
factor\ $\sim\ {1\over\tau_c^{2-A_I}}$.

\section{Discussion}

We have studied spacetimes with null cosmological singularities and
string propagation in their background: these are essentially
anisotropic plane waves with singularities. As we have discussed
(argued in \cite{knNullws2}), the free string lightcone Schrodinger
wavefunctional is regular in the vicinity of the singularity in the
Rosen frame where the null Kasner exponents satisfy $|A_I|\leq 1$,
while ill-defined in the Brinkman and other Rosen frames. Only certain
singularities admit a Rosen frame of this sort with a well-defined
wavefunctional, as we have shown by analysing the
$A_I,\chi_I$-relations between the Kasner exponents and the plane wave
parameters.  Although the wavefunctional is well-defined in these
Rosen frames suggesting well-defined time evolution across the
singularity, various physical observables for the free string, in
particular the energy, are still divergent in this Rosen frame as we
have seen. This makes the significance of the apparent regularity of
the Schrodinger wavefunctional in the ``nice'' Rosen frames less clear
and suggests that in fact the free string limit is singular in any
frame. Finally we note that various single string oscillator states
become light in the vicinity of the singularity, perhaps consistent
with the fact that the free string limit is breaking down, suggesting
that string interactions become important in the vicinity of the
singularity.  Perhaps a second quantized description, say string field
theory, is the appropriate framework for investigating if string
evolution is smooth across these singularities.

Perhaps it is worth noting that the wavefunctions discussed here are
not Hamiltonian eigenstates. What we have done is to construct a
cutoff constant-(null)time surface in the vicinity of the singularity
(akin to a stretched horizon outside a black hole), identify mode
asymptotics and thereby construct ``near-singularity'' stringy states
living on this cutoff surface. This is somewhat different in spirit
from following the time evolution towards the singularity of a given
eigenstate. Likewise the oscillator masses are the ``instantaneous''
masses, perhaps best interpreted as arising from the spectrum of
string fluctuations in the vicinity of the singularity. It would be
interesting to develop a deeper understanding of these
near-singularity string states/wavefunctionals. Part of our motivation
here stems from intuition arising from the investigations of
cosmological singularities in AdS/CFT \cite{dmnt,adnt,adnnt}. The dual
gauge theory effective action is subject to renormalization effects:
defining this precisely in a Wilsonian fashion by say constructing the
effective potential on a cutoff constant-time surface in the vicinity
of the singularity might be dual to understanding stringy effects on a
corresponding cutoff surface in the bulk. The present case of purely
gravitational plane wave spacetimes is entirely closed string, but
presumably the bulk intuition should hold nevertheless.

The spacetimes considered here can be thought of near singularity limits 
of spacetimes of the general form\ \
$ds^2 = -2dx^+dx^- + e^{f_I(x^+)} (dx^I)^2$,\ \
where the $e^{f_I}$ are null scale factors which crunch at some 
location say $x^+=0$, possibly as\ $e^{f_I}\ra (x^+)^{A_I}$. These are 
in Rosen form: the redefinition 
$y^I=e^{f_I/2} x^I ,\ y^-=x^-+{f_I' (y^I)^2\over 4}$,\ transforms this to 
a Brinkman-like form\ \
$ds^2 = -2dx^+dy^- + (dy^I)^2 + 
{1\over 4} \left((f_I')^2 + 2f_I''\right) (y^I)^2 (dx^+)^2$\ \
(with $f_I'\equiv \del_+f_I$).\ 
The Rosen form spacetimes are solutions if $R_{++}=0$, giving\ 
$\sum_I \left( {1\over 2} (f_I')^2 + f_I'' \right) = 0$ ,
which is one equation for $D-2$ scale factors $f_I$, for a $D$-dim 
spacetime. Thus the space of such cosmologies is large and we can 
choose the scale factors in various ways, to suit the physical question 
we are interested in. In particular, choosing the $e^{f_I}\ra 1$ at 
early times $x^+\ra -\infty$ renders the spacetimes asymptotically 
flat at early times, in both Rosen and Brinkman frames.
Note also that spacetimes with $e^{f_I}\ra (x^+)^{A_I}$ near 
$x^+=0$ acquire a Rosen-like null Kasner cosmology form we have 
elaborated on earlier, with the corresponding Brinkman form discussed 
earlier. Looking more closely at the null Kasner solutions, we note 
that the equation of motion can be written as\ 
$2 \sum_I A_I = \sum_I A_I^2$,\ 
very similar to the condition $\sum_ip_i=\sum_ip_i^2$ for the usual 
Kasner cosmologies with exponents $p_i$. However while the $p_i$ there 
also were required to satisfy\ $\sum_ip_i=1$ (from the two equations 
for $R_{tt},R_{ii}$), the single equation of motion here stemming from 
$R_{++}$ gives just one condition on the $A_I$, and more freedom 
in the space of such cosmologies. It might be interesting to 
understand more general null cosmologies where the spatial slices are 
curved: these would be null analogs of the well-known BKL cosmologies 
(discussed in the AdS/CFT context in \cite{adnnt}), and it would be 
interesting to explore the role of spatial curvatures, and the 
approach to the singularity.

\vspace{3mm}
\noindent {\small {\bf Acknowledgments:} It is a great pleasure to 
thank S.~Das, S.~Trivedi and especially M.~Blau and M.~O' Loughlin 
for discussions. I thank the Theory group, TIFR, Mumbai and the 
Organizers of the Ascona string theory Workshop, Switzerland, for 
hospitality during the early stages of this work. This work is 
partially supported by a Ramanujan Fellowship, DST, Govt of India.}


\end{document}